\documentclass[a4paper]{jpconf}
\usepackage{graphicx}
\usepackage{amsmath}
\usepackage{url}
\usepackage{color}

\usepackage[normalem]{ulem}  % \sout{old text} for strikeout
\renewcommand\sout{\bgroup \color{green} \ULdepth=-.5ex \ULset}
\begin{document}
\title{$K^-d\rightarrow \pi\Sigma n$ reactions and structure of the $\Lambda(1405)$}

\author{S~Ohnishi$^{1,2}$,  Y~Ikeda$^{2}$, T~Hyodo$^{3}$, E~Hiyama$^{2}$,
W~Weise$^{4,5}$}

\address{$^{1}$ Department of Physics, Tokyo Institute of Technology,
Tokyo 152-8551, Japan}
\address{
$^{2}$ RIKEN Nishina Center, Wako, Saitama 351-0198, Japan}
\address{
$^{3}$ Yukawa Institute for Theoretical Physics, Kyoto University, Kyoto 606-8502, Japan}
\address{
$^{4}$ ECT*, Villa Tambosi, I-38123 Villazzano (Trento), Italy}
\address{
$^{5}$ Physik Department, Technische Universit\"{a}t M\"{u}nchen, D-85747 Garching, Germany}

\ead{s\_ohnishi@riken.jp}

\begin{abstract}
We report on the first results of a full three-body calculation of the 
$\bar{K}NN$-$\pi YN$ amplitude for the $K^-d\rightarrow\pi\Sigma n$ reaction,
and examine how the $\Lambda(1405)$ resonance manifests itself
in the %\sout{$\pi\Sigma$ invariant-mass} 
{neutron energy} distributions of
$K^-d\rightarrow\pi\Sigma n$ reactions.
The amplitudes are computed using the 
$\bar{K}NN$-$\pi YN$ coupled-channels Alt-Grassberger-Sandhas (AGS) equations.
Two types of models are considered for the two-body meson-baryon interactions: 
an energy-independent interaction and an energy-dependent one, 
both derived from the leading order chiral SU(3) Lagrangian. These two models have
different off-shell properties that cause correspondingly different behaviors in the 
three-body system. 
As a remarkable result of this investigation, it is found that the
 %\sout{$\pi\Sigma$ invariant mass} 
{neutron energy} spectrum,
reflecting the $\Lambda(1405)$ mass distribution and width, depends 
quite sensitively on the (energy-dependent or energy-independent) model
 used. Hence accurate
measurements of the $\pi\Sigma$ mass distribution have the potential to
 discriminate between 
possible mechanisms at work in the formation of the $\Lambda(1405)$. 
\end{abstract}

\section{Introduction}
Understanding the structure of the $\Lambda(1405)$ with spin-parity
$J^\pi=1/2^-$ and strangeness $S=-1$ is a
long-standing issue in hadron physics.
The mass of the $\Lambda(1405)$ is slightly less than the $\bar{K}N$ threshold
energy.  The $\Lambda(1405)$ can be considered as a $\bar{K}N$ quasi-bound
state embedded in the $\pi\Sigma$ continuum~\cite{Dalitz:1959dn,Dalitz:1960du}.
Guided by this picture, $\bar{K}N$ interactions which reproduce the mass
of $\Lambda(1405)$ and two-body scattering data have been constructed phenomenologically~\cite{Akaishi:2002bg,Shevchenko:2011ce}. 
On the other hand, $\bar{K}N$ interactions have been studied for a
long time based on
chiral SU(3) dynamics~\cite{Kaiser:1995eg,Oset:1997it,Hyodo:2011ur}.
Between the phenomenological and chiral SU(3) $\bar{K}N$ interactions, 
subthreshold $\bar{K}N$ amplitudes are quite
different~\cite{Hyodo:2007jq}. 
The phenomenological model describes $\Lambda(1405)$ as a single
pole of the scattering amplitude around 1405~MeV.
The $\bar{K}N$ amplitude from the interaction based on chiral SU(3)
dynamics has two poles, one of which located not at 1405 MeV but around
1420 MeV~\cite{Oller:2000fj,Jido:2003cb}.
The differences in the pole structure come from the different
off-shell behavior,
especially as a consequence of the energy-dependence of the $\bar{K}N$ interaction.
The $\bar{K}N$ interaction based on chiral SU(3) dynamics
is energy-dependent, and its attraction becomes weaker as one moves below the $\bar{K}N$
threshold energy. Hence the (upper) pole of the $\bar{K}N$ amplitude shows up around 1420 MeV.
On the other hand, the phenomenological $\bar{K}N$ interaction is energy-independent and
strongly attractive so that the pole shows up around 1405 MeV.
These differences are enhanced in the so-called few-body kaonic nuclei, such as the strange
dibaryon resonance under discussion in the
$\bar{K}NN$-$\pi YN$ coupled
system~\cite{Yamazaki:2002uh, Yamazaki:2007cs, Dote:2008in, Dote:2008hw,
Wycech:2008wf, Barnea:2012qa, Shevchenko:2006xy, Shevchenko:2007ke,
Ikeda:2007nz, Ikeda:2008ub, Ikeda:2010tk}.
How a possible signature of this strange
dibaryon resonance shows up
in the resonance production reaction is also of interest as it reflects the two-body
dynamics of the $\Lambda(1405)$~\cite{Ohnishi:2013rix}.
 
One of the possible kaon-induced processes forming the $\Lambda(1405)$
is $K^-d\rightarrow \Lambda(1405)\,n$.
The signature of the
$\Lambda(1405)$ was observed in an old bubble-chamber experiment 
that measured the $\pi\Sigma$ invariant mass distribution in the
$K^-d\rightarrow \pi^+\Sigma^-n$ reaction\,\cite{Braun:1977wd}.
A new experiment is planned at J-PARC\,\cite{Noumi}. 
Theoretical investigations of the $K^-d\rightarrow \pi\Sigma n$ reaction
have previously 
been performed in simplified models assuming a two-step process\,\cite{Jido:2009jf,Miyagawa:2012xz,Jido:2012cy,YamagataSekihara:2012yv}.

In this contribution we examine how the
$\Lambda(1405)$ resonance shows up in the $K^-d\rightarrow \pi\Sigma n$
reaction by making use of the approach based on the coupled-channels
Alt-Grassberger-Sandhas~(AGS) equations developed in
Refs.~\cite{Ikeda:2007nz, Ikeda:2008ub, Ikeda:2010tk, Ohnishi:2013rix}.
This is the first calculation of this process which incorporates
the full three-body dynamics.

\section{Three-body Scattering Equations}
\label{sec:1}
Throughout this paper, we assume that the three-body processes take place via
separable two-body interactions, which have the following form 
in the two-body center-of-mass (CM) frame,
\begin{align}
V_{\alpha\beta}(\vec q_\alpha,\vec q_\beta; E) = 
g_\alpha(\vec q_\alpha) \lambda_{\alpha\beta}(E) g_\beta (\vec q_\beta) ~,
\label{eq:v_sepa}
\end{align}
where $\vec q_\alpha$ [$g_\alpha(\vec q_\alpha)$] is the relative
momentum [form factor]
of the two-body channel $\alpha$; $E$ is the total energy of the
two-body system.
With this assumption the amplitudes for the
quasi-two-body scattering of an ``isobar'' and a spectator particle, $X_{ij}(\vec p_i, \vec p_j; W)$, are then obtained by solving
the AGS equations~\cite{Alt:1967fx,PhysRev.132.485},
\begin{align}
	X_{ij}({\vec p}_i,{\vec p}_j,W)&=(1-\delta_{ij})Z_{ij}({\vec p}_i,{\vec p}_j,W) 
\nonumber\\
&
	+\sum_{n\ne i}\int d{\vec p}_n Z_{in}({\vec p}_i,{\vec p}_n,W)
	\tau_n\left(W-E_n(\vec p_n)\right) X_{nj}({\vec p}_n,{\vec p}_j,W)~.
	\label{AGS}
\end{align}
Here the subscripts $i,j,n$ specify the reaction channels; $W$ and $\vec p_i$ are the total scattering energy and the relative momentum of channel $i$ 
in the three-body CM frame, respectively;
$Z_{ij}({\vec p}_i,{\vec p}_j;W)$ and $\tau
_i\left(W-E_i(\vec p_i)\right)$ are the one-particle exchange potential and
the two-body propagator.
% given by
% %
% \begin{align}
%  Z_{ij}(\vec p_i, \vec p_j; W) &= 
% \frac{g_i(\vec q_i)g_j^\ast(\vec q_j)}{W - E_i(\vec p_i) - E_j(\vec p_j) - E_k (\vec p_k) + i\epsilon}~,\\
% \label{eq:z-diagram}
% 	\tau_i(W-E_i(\vec p_i)) &= \left[ 1/\lambda_i-\int q_i^2d{ q}_i
% 		\frac{|g_i({ q}_i)|^2}
% 		{W-E_i({\vec p}_i)-E_{jk}({\vec p}_i,{\vec q}_i)}\right] ^{-1}~~.
% \end{align}

With the quasi-two-body amplitudes,
the scattering amplitudes for the break-up process $d+\bar{K}\rightarrow \pi
+\Sigma+ N$ 
are obtained as 
\begin{align}
T_{\pi\Sigma N\text{-}\bar{K}d}(\vec q_N,\vec p_N, \vec p_{\bar{K}},W)
&=
g_{Y_\pi}(\vec q_N) \tau_{Y_\pi Y_K}\left(W-E_N(\vec p_N)\right) X_{Y_K d}(\vec p_N, \vec p_{\bar{K}},W)
\nonumber\\
&+
 g_{Y_\pi}(\vec q_N) \tau_{Y_\pi Y_\pi}\left(W-E_N(\vec p_N)\right) X_{Y_\pi d}( \vec p_N, \vec p_{\bar{K}},W)
\nonumber\\
&+
g_{N^*}(\vec q_\Sigma) \tau_{N^*N^*}\left(W-E_\Sigma(\vec p_\Sigma)\right) X_{N^* d}( \vec p_\Sigma, \vec p_{\bar{K}},W)
\nonumber\\
&+
g_{d_y}(\vec q_\pi) \tau_{d_yd_y}\left(W-E_\pi(\vec p_\pi)\right) X_{d_y d}( \vec p_\pi, \vec p_{\bar{K}},W)~,
\label{eq:t_break}
\end{align}
where $X_{Y_K d}(\vec p_N, \vec p_{\bar{K}},W)$ is the quasi-two-body
amplitude anti-symmetrized for two nucleons; the subscripts denote the
isobars.
The notations for the isobars are $Y_K=\bar{K}N$,
$Y_\pi = \pi Y$, $d = NN$, $N^*=\pi N$ and $d_y=YN$, respectively.

In this contribution we employ the first two terms of 
Eq.~(\ref{eq:t_break}) as a first step.
These terms emerge directly from the $\Lambda(1405)$ in the final
state interaction; they are the dominant parts of the full T-matrix. 
Using this T-matrix, the differential cross section of the
break-up process  
$d+\bar{K}\rightarrow \pi +\Sigma+ N$ is calculated as:
\begin{align}
 % \frac{d^2\sigma}{dM_{\pi\Sigma}d(\cos\theta_{p_N})} &=
 % (2\pi)^4\frac{E_dE_{\bar{K}}}{Wp_{\bar{K}}}\frac{m_N m_\Sigma
 % m_\pi}{m_N+m_\Sigma+m_\pi} \nonumber\\
 % &\times\int d\phi_{p_N}d\Omega_{q_N}
 % p_Nq_N \sum_{\bar{i}f}|<N\Sigma \pi|T(W)|d\bar{K}>|^2~~,\label{eq:differential}\\
 \frac{d\sigma}{d{E_n}} &= (2\pi)^4\frac{E_dE_{\bar{K}}}{Wp_{\bar{K}}}\frac{m_N m_\Sigma
 m_\pi}{m_N+m_\Sigma+m_\pi}\nonumber\\
 &\times\int d\Omega_{p_N}d\Omega_{q_N}
 p_Nq_N \sum_{\bar{i}f}|<N\Sigma \pi|T(W)|d\bar{K}>|^2~~\label{eq:differential2},
\end{align}
where %\sout{$M_{\pi\Sigma}$ is the non-relativistic invariant/missing mass}
$E_n$ is the neutron energy in the center-of-mass frame of $\pi\Sigma$
defined by
\begin{align}
 E_n  =  m_N  + \frac{p_N^2}{2\eta_N} ~~.\label{eq:inv_mass}
\end{align}

\section{Models of Two-body Interaction}
\label{sec:4}
We use two-body $s$-wave meson-baryon interactions obtained from 
the leading order chiral Lagrangian,
\begin{equation}
L_{WT} = \frac{i}{8F_\pi^2}
  Tr(\bar{\psi}_B\gamma^\mu[[\phi,\partial_\mu\phi],\psi_B]).
\end{equation}
Here, we examine two interaction
models, both of which are derived from the above Lagrangian  
but have different off-shell behavior:
one is the energy dependent model (E-dep),
\begin{eqnarray}
V_{\alpha \beta}(q',q;E)
=&-&\lambda_{\alpha\beta}\frac{1}{32\pi^2 F_\pi^2}
\frac{2E-M_\alpha -M_\beta}{\sqrt{m_\alpha m_\beta}}
\left( \frac{\Lambda_\alpha^2}{q'\,^2+\Lambda_\alpha^2} \right)^2
\left( \frac{\Lambda_\beta^2}{q^2+\Lambda_\beta^2} \right)^2.
\label{eq:e-dep}
\end{eqnarray}
while the other is the energy independent model (E-indep),
\begin{eqnarray}
V_{\alpha \beta}(q',q)
=&-&\lambda_{\alpha\beta}\frac{1}{32\pi^2 F_\pi^2}
\frac{m_\alpha +m_\beta}{\sqrt{m_\alpha m_\beta}}
\left( \frac{\Lambda_\alpha^2}{q'\,^2+\Lambda_\alpha^2} \right)^2
\left( \frac{\Lambda_\beta^2}{q^2+\Lambda_\beta^2} \right)^2 ,
\label{eq:e-indep}
\end{eqnarray}
Here, $m_\alpha$ ($M_\alpha$) is 
the meson (baryon) mass; 
$F_\pi$ is the pion decay constant; 
$\lambda_{\alpha\beta}$ are determined by the 
flavor SU(3) structure of the chiral Lagrangian. 

In the derivation of these potentials we have assumed
the so-called ``on-shell factorization''~\cite{Oset:1997it}
for Eq.~(\ref{eq:e-dep}) and
$q, q'\ll M_\alpha$ for Eq.~(\ref{eq:e-indep}).
The cutoff parameters $\Lambda$ are determined by fitting 
experimental data {as shown in} Table~\ref{tab:1}.

In the E-dep model, the $\bar{K}N$ amplitudes have two poles for $l=I=0$
{in the $\bar{K}N$-physical and $\pi\Sigma$-unphysical sheets},
corresponding to 
those derived from the chiral unitary model~\cite{Jido:2003cb}. 
On the other hand, the E-indep model has a single pole 
that corresponds to $\Lambda(1405)$.
It is interesting to examine how 
this difference of the two-body interaction models 
appears in the %\sout{invariant mass distribution} 
neutron energy spectrum of the $K^- d \rightarrow
\pi\Sigma n$ reaction.

\begin{table}
\caption{Cutoff parameters of $\bar{K}N$-$\pi Y$ interaction.}
\centering
\label{tab:1}       % Give a unique label
\begin{tabular}{lccccc}
\br
 &$\Lambda ^{I=0}_{\bar{K}N}$(MeV) &$\Lambda ^{I=0}_{\pi\Sigma}$(MeV)
 &$\Lambda ^{I=1}_{\bar{K}N}$(MeV) &$\Lambda
 ^{I=1}_{\pi\Sigma}$(MeV)&$\Lambda ^{I=1}_{\pi\Lambda}$(MeV) \\[3pt]
%\tableheadseprule\noalign{\smallskip}
\mr
E-dep&1000 & 700 & 725&725&725 \\
E-indep&1000 & 700 & 920&960&640 \\
\br
\end{tabular}
\end{table}

\section{Results and Discussion}
\label{sec:5}

\begin{figure}
   \includegraphics[width=0.5\textwidth]{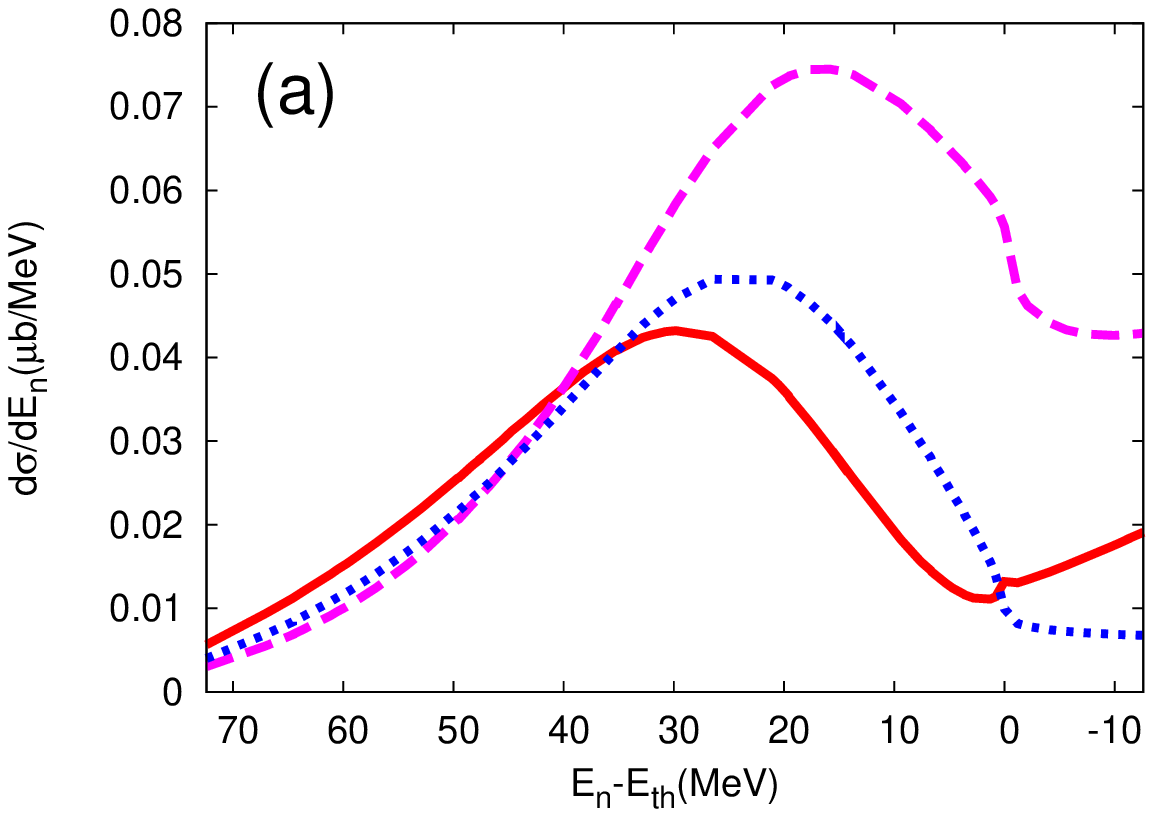}
   \includegraphics[width=0.5\textwidth]{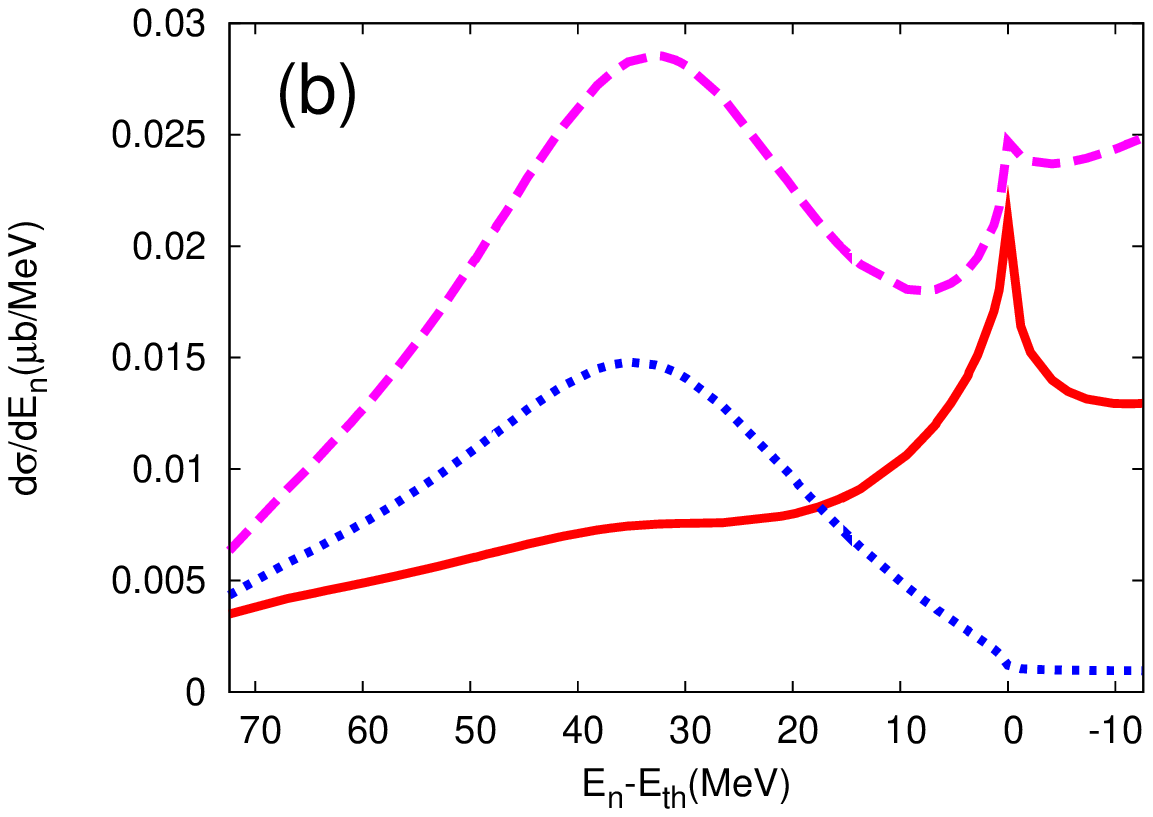}
 \caption{Differential cross section %\sout{$d\sigma/dM_{\pi\Sigma}$}
 {$d\sigma/dE_n$} 
 for $K^- + d\rightarrow \pi+\Sigma +n$.
 {The initial kaon momentum is set to $p_{K^-}^{lab}=1000$~MeV.}
 Panel~(a): the E-dep model; Panel~(b) the E-indep model.
 Solid curves: $\pi^+\Sigma^-n$;
dashed curves: $\pi^-\Sigma^+n$;
dotted curves: $\pi^0\Sigma^0n$ in the final
state, respectively.
 }
 \label{fig:1}
\end{figure}

In Fig.\ref{fig:1}, we present the differential cross section
of $K^- d\rightarrow \pi \Sigma n$ [Eq. (\ref{eq:differential2})]
computed using the E-dep (a) and E-indep (b) models, respectively.
We investigate the cross section for initial kaon momentum
$p_{K^-}^{lab}=1000$ MeV {in accordance with the planned J-PARC
experiment}~\cite{Noumi}.
Here, we decompose the isospin basis states into charge basis states
using Clebsch–Gordan coefficients:
the solid curve represents the $K^- + d\rightarrow\pi^++\Sigma^-+n$;
the dashed curve refers to the $K^- + d\rightarrow\pi^-+\Sigma^++n$;
the dotted curve represents the $K^- + d\rightarrow\pi^0+\Sigma^0+n$
reaction, respectively.
% \sout{
% We observe the $\bar{K}N$ threshold cusp not at $M_{\pi\Sigma}\sim 1433$ MeV
% but at $M_{\pi\Sigma}\sim 1408$ MeV, because the
% relativistic correction for invariant/missing mass $M_{\pi\Sigma}$
% defined by Eq.~(\ref{eq:inv_mass}) is not small for the initial kaon momentum
% $p_{K^-}^{lab}=1$ GeV.}

We subtract the neutron energy $E_{th}$ at which the amplitudes have the
$\bar{K}N$ threshold cusp from the neutron energy $E_n$, 
i.e. $\bar{K}N$ threshold cusp shows up on
the differential cross section at $E_n-E_{th} = 0$.
Well defined maxima are found at %\sout{$M_{\pi\Sigma}\sim 1380$-$1390$ MeV} 
$E_n\sim 17$-$30$ MeV for the E-dep model and 
a peak or bumps at %\sout{$M_{\pi\Sigma}\sim 1370$-$1375$ MeV} 
{$E_n\sim 32$-$38$ MeV} for the E-indep model, depending in the charge
combination of $\pi\Sigma$ in the final state.
These peak and bump structures appear about %\sout{35 MeV lower} 
5 MeV higher in energy than the calculated
%\sout{resonance} 
binding energy of the
$\Lambda(1405)$ (%\sout{$E_R\sim 1420$ MeV} 
$E_B\sim 13$ MeV for the E-dep model and
%\sout{$E_R\sim1405$ MeV} 
$E_B\sim 28$ MeV  for the E-indep model). 
The magnitude of the differential cross section for the E-dep model is
twice larger than that for the E-indep model, and the interference
patterns with backgrounds are quite different between these two models.
This clear difference in the differential cross section,
arising from the model dependence of the two-body interactions,
suggests that the $K^- d\rightarrow \pi \Sigma n$ reaction can indeed provide
useful information on the $\bar{K}N$-$\pi Y$ system.

\begin{figure}
   \includegraphics[width=0.5\textwidth]{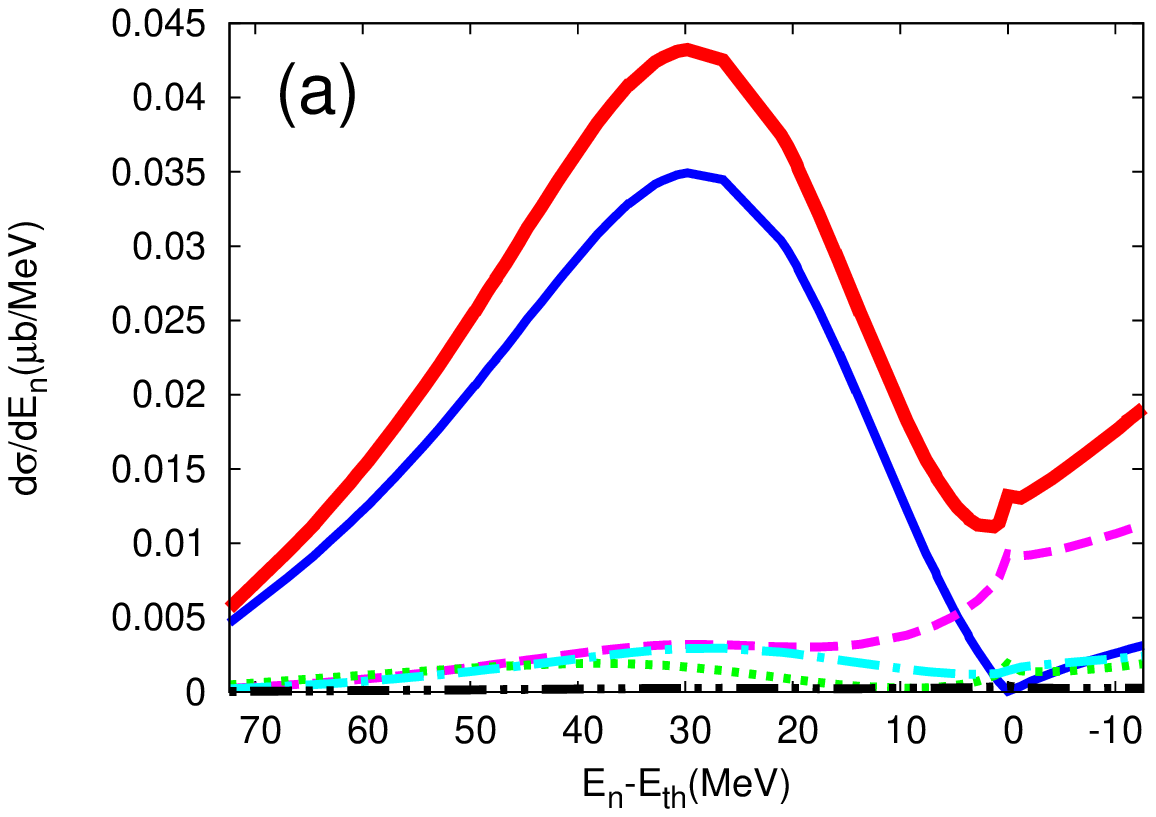}
   \includegraphics[width=0.5\textwidth]{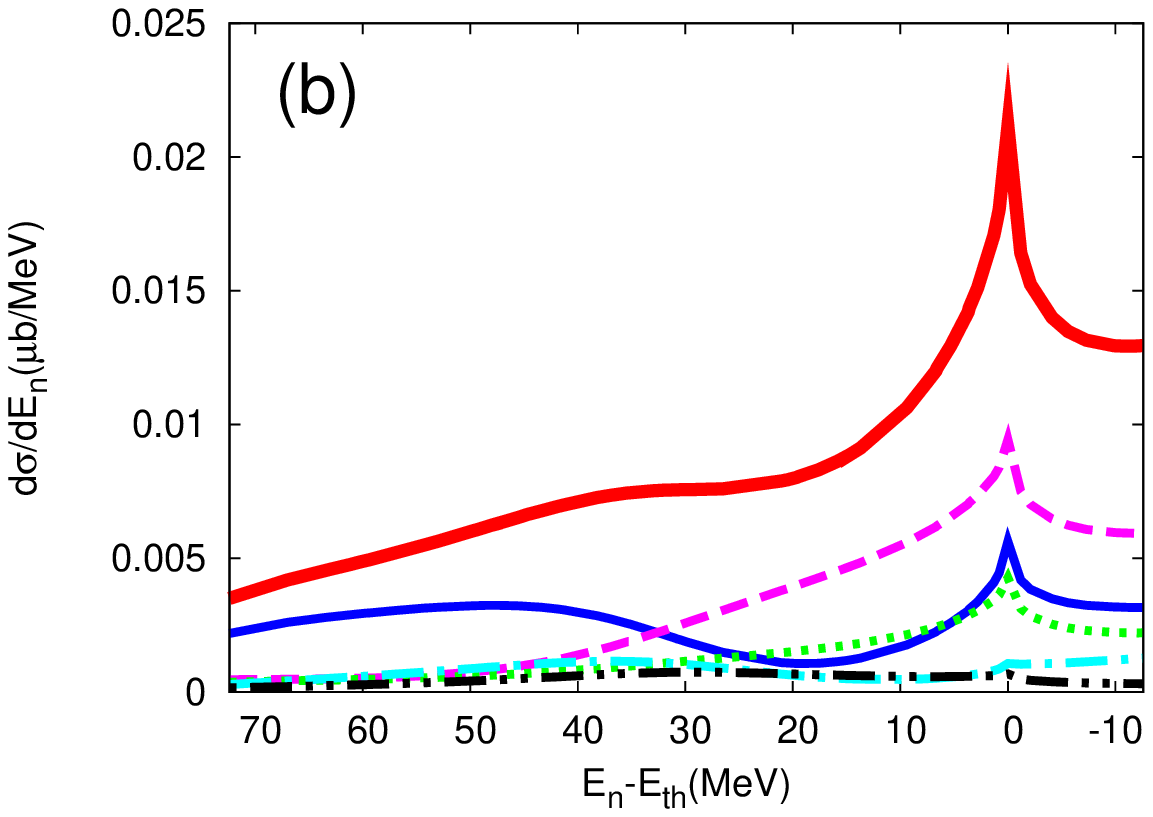}
 \caption{
Contribution of each partial wave component to the differential cross
 section %\sout{$d\sigma/dM_{\pi\Sigma}$} 
 {$d\sigma/dE_n$} 
 for $d+ K^-\rightarrow \pi^++\Sigma^- +n$.
 Panel~(a): the E-dep model; Panel~(b) the E-indep model.
 The thick solid curve represents the summation of total orbital angular
 momentum $L=0$ to $14$;
 The thin solid curve represents $L=0$ only;
 The dashed curve represents $L=1$ only;
 The dotted curve represents $L=2$ only;
 The dashed-dotted curve represents $L=3$ only;
 The dashed-two-dotted curve represents $L=4$ only, respectively.
 The initial kaon momentum is set to $p_{K^-}^{lab}=1000$~MeV.}
 \label{fig:2}
\end{figure}

Finally, we examine the contribution of each partial wave component for
total orbital angular momentum $L$ to the
differential cross section (Fig.~\ref{fig:2}).
We conclude that the $s$-wave component is dominant in the low-energy region,
but around the $\bar{K}N$ threshold higher partial waves such as the
$p$-wave component become important.

In summary, we have calculated the differential cross sections
(\ref{eq:differential2})
for $K^- + d \rightarrow \pi+\Sigma+ n$ reactions.
We have found peak and bump structures
in the %\sout{$\pi\Sigma$} 
neutron energy spectrum, and therefore it is possible to
observe the signal of the
$\Lambda(1405)$ resonance in the physical cross sections.
We have also shown that the $K^- d\rightarrow \pi\Sigma n$ reactions are
useful for judging existing dynamical models of $\bar{K}N$-$\pi\Sigma$
coupled systems with $\Lambda(1405)$. 
% In the current work, however, we have not taken into account the
% contribution from third
% and forth term of the T-matrix~(\ref{eq:t_break}) and relativistic correction. 
% Further improvements of the current model and the calculation of the 
% actual cross sections are under investigation.
{Further improvements of the present model to account for the
neglected contributions in Eq.~(\ref{eq:t_break}) and relativistic
corrections are under investigation.}

\ack
The simulation has been performed on a supercomputer (NEC SX8R) at the Research Center
for Nuclear Physics, Osaka University.
This work was partly supported by RIKEN Junior Research Associate Program, 
by RIKEN iTHES Project
{and by JSPS KAKENHI Grants Nos. 25800170, 24740152 and 23224006}.

\section*{References}
\bibliography{sotancp3}
%\begin{thebibliography}{9}
% \bibitem{iopartnum} IOP Publishing is to grateful Mark A Caprio, Center for Theoretical Physics, Yale University, for permission to include the {\tt iopart-num} \BibTeX package (version 2.0, December 21, 2006) with  this documentation. Updates and new releases of {\tt iopart-num} can be found on \verb"www.ctan.org" (CTAN). 

% \end{thebibliography}

\end{document}